# Statistical Decision Properties of Imprecise Trials Assessing COVID-19 Drugs

Charles F. Manski
Department of Economics and Institute for Policy Research, Northwestern University

and

Aleksey Tetenov
Geneva School of Economics and Management, University of Geneva

May 30, 2020

## Abstract

As the COVID-19 pandemic progresses, researchers are reporting findings of randomized trials comparing *standard care* with care augmented by experimental drugs. The trials have small sample sizes, so estimates of treatment effects are imprecise. Seeing imprecision, clinicians reading research articles may find it difficult to decide when to treat patients with experimental drugs. Whatever decision criterion one uses, there is always some probability that random variation in trial outcomes will lead to prescribing sub-optimal treatments. A conventional practice when comparing standard care and an innovation is to choose the innovation only if the estimated treatment effect is positive and statistically significant. This practice defers to standard care as the status quo. To evaluate decision criteria, we use the concept of *near optimality,* which jointly considers the probability and magnitude of decision errors. An appealing decision criterion from this perspective is the *empirical success* rule, which chooses the treatment with the highest observed average patient outcome in the trial. Considering the design of recent and ongoing COVID-19 trials, we show that the empirical success rule yields treatment results that are much closer to optimal than those generated by prevailing decision criteria based on hypothesis tests.

We have benefitted from the comments of Michael Gmeiner, Valentyn Litvin, Francesca Molinari, and John Mullahy. Tetenov has received funding from the Swiss National Science Foundation through grant number 100018-192580.

1. Introduction

As the COVID-19 pandemic progresses, researchers are reporting findings of randomized trials comparing standard care with care augmented by experimental drugs. The trials to date have small sample sizes, so estimates of treatment effects are statistically imprecise. Seeing imprecision, clinicians reading research articles may find it difficult to decide when to treat patients with experimental drugs. Whatever decision criterion one uses, there is always some probability that random variation in trial outcomes will lead to prescribing sub-optimal treatments.

A conventional practice when comparing standard care and an innovation is to choose the innovation only if the estimated treatment effect is positive and statistically significant. This practice defers to standard care as the status quo. To evaluate decision criteria, we use the concept of *near optimality,* which jointly considers the probability and magnitude of decision errors. An appealing decision criterion from this perspective is the *empirical success* rule, which chooses the treatment with the highest observed average patient outcome in the trial.

The contributions of this paper are both applied and methodological. We apply to recent COVID-19 trials the methodology for study of two-arm trials developed earlier in Manski (2004) and Manski and Tetenov (2016, 2019). We extend the computational reach of the methodology to enable practical analysis of multi-arm trials, such as the ongoing Recovery Trial in the United Kingdom. We show that the empirical success rule yields treatment results that are much closer to optimal than those generated by prevailing decision criteria based on hypothesis tests.

2. Background

A core objective of randomized trials is to inform treatment choice. When comparing standard care with an innovation, the prevailing statistical practice has been for researchers to conclude that the innovation is better than standard care only if the estimated average treatment effect comparing the innovation with standard care is statistically significant by conventional criteria. Equivalently, a conventional statistical hypothesis test must reject the null hypothesis that the innovation is no better than standard care.

Statistical analysis commonly examines each pre-declared primary and secondary outcome of a trial in isolation from one another rather than the joint effect of all outcomes. Articles reporting trials often report subgroup findings only when they are statistically significant. They often discuss side effects of treatments without performing statistical analysis. We further consider these practices in Section 5.

To illustrate, Box 1 summarizes a trial comparing standard-care treatment for severe COVID-19 with standard care augmented by prescription of lopinavir–ritonavir. A clinician might reasonably view the estimated reductions in median time to clinical improvement and in mortality to be suggestive, albeit not definitive, evidence that treatment with lopinavir–ritonavir is beneficial relative to standard care alone. Yet the study authors conclude (p. 1) "no benefit was observed with lopinavir–ritonavir treatment beyond standard care." This conclusion was reached because the estimated treatment effects were not statistically significant, having confidence intervals that cover zero. Subsequently, COVID-19 treatment guidelines issued by the National Institute of Health (NIH) cited the absence of statistical significance when it characterized the study as having negative findings.[1]

---

[1] The NIH document states: "The Panel **recommends against** the use of **lopinavir/ritonavir (AI) or other HIV protease inhibitors (AIII)** for the treatment of COVID-19, except in the context of a clinical trial. It later remarks: "There was a lower, but not statistically significant, mortality rate (lopinavir/ritonavir 19.2%,

Requiring statistical significance to prescribe a treatment an innovation shows deference to standard practice, placing the burden of proof on the innovation. From a clinical perspective, one might argue that it is reasonable to place the burden on an innovation when standard care is known to yield good patient outcomes. However, this argument lacks appeal in the COVID-19 setting. Current standard practice for treating COVID-19 has developed rapidly to cope with an emergency. It has not been shown to yield notably good patient outcomes. How might clinicians act with imprecise evidence such as in the Cao *et al.* study?

The statistical analysis of the Cao *et al.* study is not unusual. A much-anticipated preliminary report on a large trial comparing standard care augmented by the drug remdesivir with standard care alone concludes that remdesivir reduces time to recovery, but the report states no conclusion regarding mortality (Beigel *et al.* 2020). The report states that remdesivir reduces median time to recovery from 15 days to 11 days and reduces 14-day mortality from 11.9% to 7.1%. A clinician might reasonably consider both findings to be relevant to treatment choice, but the research team stated a conclusion only for time to recovery. The rationale was that the former finding was statistically significant by conventional criteria and the latter was insignificant.[2]

Consider the major new Recovery Trial under development in the United Kingdom (https://www.recoverytrial.net/). This trial will compare standard care alone with standard care augmented by various experimental drugs. The statistical analysis plan calls for application of hypothesis tests to compare patient outcomes under alternative treatments.

---

on SOC 25.0%) and shorter ICU stay compared to those given SOC (6 days vs. 11 days; difference = -5 days; 95% CI, -9 to 0)."
https://covid19treatmentguidelines.nih.gov/therapeutic-options-under-investigation/antiviral-therapy/, accessed May 24, 2020.

[2] The reported 95% confidence intervals for the rate ratio for recovery and the hazard rate for death were [1.12, 1.55] and [0.47, 1.04] respectively. The former interval is deemed a statistically significant treatment effect by conventional criteria because the lower bound 1.12 is slightly larger than 1. The latter interval is deemed an insignificant treatment effect because the upper bound 1.04 is slightly larger than 1.

---

Box 1: Statistical Analysis in a Trial Comparing Treatments for COVID-19

Cao *et al.* (2020) report on a randomized trial in China comparing standard-care treatment of severe cases of COVID-19 with standard-care combined with the drug pair lopinavir–ritonavir. The trial assigned 99 hospitalized adult patients to the lopinavir–ritonavir group and 100 to the standard-care only group. The pre-declared primary endpoint measured time to clinical improvement. A secondary outcome was mortality within 28 days.

The authors summarized the primary finding as follows (p. 1): "In a modified intention-to-treat analysis, lopinavir–ritonavir led to a median time to clinical improvement that was shorter by 1 day than that observed with standard care (hazard ratio, 1.39; 95% CI, 1.00 to 1.91). Regarding mortality, 19 of the 99 patients assigned to lopinavir-ritonavir died within 28 days and 25 of the 100 receiving only standard care died. The authors characterized this finding as follows (p. 1): "Mortality at 28 days was similar in the lopinavir–ritonavir group and the standard-care group (19.2% vs. 25.0%; difference, −5.8 percentage points; 95% CI, −17.3 to 5.7)." They reported raw findings on side effects, but they performed no statistical analysis. They concluded (p. 1): In hospitalized adult patients with severe Covid-19, no benefit was observed with lopinavir-ritonavir treatment beyond standard care.

---

3. Measuring the Near-Optimality of Criteria Using Trial Data to Choose a Treatment

The results observed in any randomized trial have random variation. Whatever criterion one uses to make treatment decisions based on the results of a trial, there is always some probability that this random variation will lead to prescribing a sub-optimal treatment to patients. Considering the probability of error alone is insufficient. The same error probability should be less tolerable when the impact of sub-optimal treatment on patient welfare is larger. To evaluate treatment choice based on trial data, we use the concept of *near-optimality* (Manski and Tetenov, 2016, 2019), which jointly considers the probability of errors and their magnitudes.

The concept is as follows. Consider specified possible values for average patient outcomes under each treatment. Presuming the common medical focus on average patient outcomes, the ideal clinical decision would be to prescribe a treatment that maximizes average outcome. Trial data do not reveal the best treatment with certainty, so one cannot achieve this ideal. Suppose then that one applies some decision criterion to the trial data. The criterion may be a hypothesis test or another one that we will introduce shortly.

For every treatment that is not best, we compute the probability that it would be prescribed when the criterion is applied to the results of a trial. We multiply this error probability by the magnitude of the loss from prescribing this treatment, measured by the difference in average patient outcomes compared to the best treatment. This product measures the expected loss from prescribing the inferior treatment, also called its *regret*. The sum of these expected losses across all inferior treatments measures the gap between the ideal of prescribing the best treatment and the reality of having to prescribe the treatment based on trial data that is subject to random variation.

The above calculations are made using specified possible values for average patient outcomes with each treatment. However, trial data do not reveal the true values for average patient outcomes; they only enable one to estimate them. The final measurement step is to look across all a priori possible values for average patient outcomes for all treatments to find the values where the expected loss from prescribing inferior treatments is largest. This measures the nearness to optimality of the proposed criterion for clinical decision making. Nearness to optimality is also called *maximum regret*. See the Technical Appendix for a mathematical statement.

To illustrate how measurement of nearness to optimality works in practice, Table 1 applies two different decision criteria to the trial design in Cao *et al.* (2020), which assigned 100 patients to standard care and 99 to care augmented by lopinavir–ritonavir. We focus on the outcome of 28-day mortality, presuming that this is the most important outcome for patients with severe COVID-19. Each column in the table specifies one scenario for average patient outcomes, combining a mortality rate of standard care, fixed at 0.25, with a mortality rate of the new treatment, ranging from 0.4 to 0.1.

Panel A shows what would happen if the trial data were used to make treatment decisions with a conventional two-sided hypothesis test at 5% level. Thus, the new treatment would be prescribed if the results of the hypothesis test show the new treatment to be statistically significantly better than standard care. If the new treatment is better, prescribing standard care is an error. The loss from this error is the difference in average patient outcomes, in this example mortality.

The table shows that if the new treatment has a mortality rate of 0.15, compared to 0.25 for the standard care, a trial with the design of Cao *et al*. will erroneously reach a negative conclusion about the new treatment in 57.4% of trials, leading clinicians to continue using standard care. The magnitude of the error is 0.1, the difference between 0.25 and 0.15. Multiplying the probability of error by its magnitude gives an expected loss of 0.0574.

Suppose instead that the new treatment has mortality rate 0.2. Then the hypothesis test would reach a negative conclusion about the new treatment in 86.8% of trials. While the error probability in this scenario is higher, it is less consequential for clinical outcomes because the difference in mortality rates between treatments is 0.05. In this case, expected loss is 0.868 x 0.05 = 0.0434.

If the new treatment has mortality rate 0.3 (0.05 higher than standard care), the hypothesis test would reach a positive conclusion about the new treatment only in 0.3% of trials, leading to expected loss of 0.003 x 0.05 = 0.00015. Expected loss is also extremely low in other scenarios where the new treatment has a considerably higher mortality rate than standard care because the probability of Type I error of a hypothesis test is dramatically lower than its nominal size. The nominal size 0.05 of the test is the error probability in the borderline case where the two treatments have the same mortality rate.[3]

We measure nearness to optimality by considering all possible scenarios for the average outcomes of treatments in the trial, which can take any values in the [0, 1] interval, not just the few scenarios illustrated in Table 1. We report nearness to optimality for treatment choice based on hypothesis tests in two-arm trials with different sample sizes in Table 2. The table shows that choosing treatments based on a hypothesis test following a two-arm trial in which 100 patients receive each treatment (as in Cao *et al.*) achieves near-optimality of 0.071. Specifically, the maximum value of expected loss across all possible values of average mortality rates occurs when the new treatment has mortality rate 0.548 and standard care has rate 0.661. Then the expected loss (0.661 − 0.548) multiplied by the error probability 0.624 equals 0.071.

Hypothesis tests by design treat standard care and the new treatment asymmetrically. An appealing alternative decision criterion is the *empirical success* rule, which we have previously studied in Manski (2004) and Manski and Tetenov (2016, 2019). This criterion chooses the treatment with the highest observed average patient outcome in the trial, regardless of the statistical significance of the result. Whereas hypothesis testing favors standard care and places the burden of proof on innovations, the empirical success rule assesses the evidence on each treatment symmetrically.

The properties of the empirical success rule are illustrated in Panel B of Table 1. If the new treatment has mortality rate 0.2 and standard care has rate 0.25, using the empirical success rule will result in prescribing the new treatment in 78.8% of trials, whereas the hypothesis testing approach of Panel A would only do so in 13.2% of trials. Given that the empirical success rule treats both treatments symmetrically, the expected losses in situations when the new treatment is better and when standard care is better are also symmetric.

---

[3] A two-sided hypothesis test rejects the null hypothesis both if the new treatment performs sufficiently better and if the new treatment performs sufficiently worse than standard care. The allowed Type I error probability is split between these two cases, but rejection of the statistical null hypothesis only leads to the prescription of the new treatment in the first case.

Table 2 compares near-optimality of the empirical success rule and the hypothesis test-based decision criterion in two-arm trials for a wide range of sample sizes. These calculations consider all possible values for the average mortality rates of the two treatments. The Technical Appendix describes the algorithm used to compute near-optimality.

The empirical success rule is about 6 times nearer to optimality than the test-based decision criterion. For example, in a trial with 100 patients in each arm (similar to Cao *et al.*), the empirical success rule achieves near-optimality of 0.012. The maximum value of expected loss occurs when standard care and the new treatment have mortality rates of 0.527 and 0.473. In this case, standard care is erroneously prescribed with probability 0.226. The empirical success rule is symmetric, so the same expected loss occurs when standard care has mortality rate 0.473 and the new treatment has rate 0.527. Then the new treatment is also erroneously prescribed with probability 0.226.

Good near-optimality properties of the empirical success rule in two-arm trials are well established in the theoretical literature. Given any specified sample size, the empirical success rule has been shown to achieve the lowest possible value of near-optimality in trials with binary outcomes that assign an equal number of patients to each arm (Stoye, 2009). It has also been shown to do so asymptotically in general trials comparing two treatments (Hirano and Porter, 2009).

What are the implications of near-optimality for clinical decision making and trial design? Suppose that a clinician were to choose between standard care and standard care augmented with lopinavir/ritonavir based solely on the results of Cao *et al.*, using standard hypothesis testing methodology. Then the average mortality rate of these patients could be up to 0.071 higher than under the better of the two treatments. (The average here is taken over different possible outcomes the trial could produce.) Given the gravity of the patient outcomes at stake, this may be an unacceptably high expected loss in welfare.

There are two ways of reducing maximum expected loss: (i) increase the sample size of the trial and (ii) change the way the trial results are translated into clinical practice. Table 2 shows that a trial enrolling 4000 patients into each arm, followed by treatment choice using standard hypothesis testing, would achieve near-optimality of 0.0115. About the same level of near-optimality (0.0120) could be achieved by using the empirical success rule in a trial with 100 patients in each arm. Thus, the empirical success rule yields a dramatic improvement in near-optimality relative to hypothesis testing.

Medical research evaluating pharmaceuticals has traditionally shown deference to standard care.[4] Hence, one might question the empirical success rule on the grounds that it evaluates the treatments in the trial symmetrically, and thus has the same levels of Type I and Type II errors. We think symmetric evaluation of standard care and innovations is justified in the COVID-19 setting when considering trials that compare carefully chosen treatments, without a financial conflict of interest, and that report all patient-relevant outcomes.

First, standard care for COVID-19 patients has not itself been perfected through clinical trials. Hence, it is unclear why it should get the "benefit of the doubt." Second, it would be unethical to participants to conduct clinical trials of a new treatment if they were ex ante expected to yield worse outcomes than standard care. This suggests an ex ante ethical symmetry between the possibilities that the new treatment is better, and that standard care is better. It is logical, then, to evaluate the two treatments symmetrically.

---

[4] The specific form of deference differs between *superiority trials* and *non-inferiority trials*. The null hypothesis used in the former trials defers to standard care somewhat more than that used in the latter trials. See U. S. Food and Drug Administration (2016).

4. Near-Optimality in Multi-Arm Trials

A number of promising pharmaceutical treatments for COVID-19 are currently undergoing clinical trials. Most of them are two-arm trials, comparing one experimental treatment with standard care. It is important for clinicians to learn not only which treatments are better than standard care, but also which treatments are the most effective.

Running multiple two-arm trials has a significant drawback when there are concurrently several treatments under investigation: the performance of alternative treatments cannot easily be compared between trials because the populations from which different trials recruit patients usually are not the same. Trials may also differ in the characteristics of standard care they provide and in the outcomes they report. These problems are addressed by multi-arm trials that randomize the same patients either to standard care or to one of several experimental treatments.

Two large-scale multi-arm trials of treatments for COVID-19 are currently under way. The Recovery Trial in the United Kingdom is comparing standard care with at least four alternatives: Lopinavir-Ritonavir, low-dose Dexamethasone, Hydroxychloroquine, and Azithromycin. The international Solidarity Trial organized by the World Health Organization is comparing standard care with Remdesivir, Chloroquine or Hydroxychloroquine, Lopinavir-Ritonavir, and Lopinavir-Ritonavir plus Interferon beta-1a. We will consider the design of the Recovery Trial, which assigns patients to standard care and alternative treatments in a 2:1:1:1:1 ratio.

The standard way to analyze the results of multi-arm trials has been to compute a t-statistic for the difference in average trial outcomes between each new treatment and standard care. Each t-statistic is then compared to a critical value adjusted for multiplicity of hypotheses. The aim of this adjustment is to guarantee that in a scenario when all new treatments have the same true average outcome as standard care, there is only a 0.05 probability that any of the differences will be found to be statistically significant in a trial. The protocol of the Recovery Trial, for example, specifies that its results will be reported this way. The protocol specifically states that Dunnett's test of multiple hypotheses will be used. The intention to use Dunnett's test may have motivated the study team to assign patients in a 2:1:1:1:1 ratio, which has been recommended when applying this test (Dunnett, 1964).

The concept of near-optimality is well-suited to interpret the findings of a multi-arm clinical trial, as it takes into account both the probability and the magnitude of different types of errors: prescribing standard care when one of the new treatments is superior, prescribing a new treatment that is inferior to standard care, or prescribing one new treatment when another new treatment is superior.

Table 3 illustrates how the near-optimality of a decision criterion is evaluated in a multi-arm trial. We consider a trial, similar in design to the Recovery Trial, randomizing 1500 patients: 500 to standard care and all others to one of four new treatments (250 to each). The table shows what happens in one specific scenario when the mortality rate of standard care is 0.25 and the mortality rates of treatments A, B, C, and D are 0.15, 0.2, 0.3, and 0.35.

Panel A shows what would happen if the trial data were used to make treatment decisions based on a two-sided Dunnett's test at 5% level. We assume that standard care will be prescribed if none of the new treatments has a lower mortality rate that is statistically significantly better. If one or more new treatments is considered statistically significantly better, then the new treatment with the lowest mortality rate among them will be prescribed.

Treatment A has the lowest mortality rate in this scenario and will be prescribed after 70.6% of trials. Standard care will be prescribed after 25.7% of trials. Since standard care has mortality rate that is 0.1 higher than the best treatment (A), this error will lead to a loss of 0.1. The expected loss from prescribing standard care is the product of the error probability and its magnitude: 0.257 x 0.1 = 0.0257. Treatment B will be prescribed after 3.8% of trials. Since its mortality rate is 0.05 higher than that of the best treatment, the expected loss from prescribing treatment B is 0.038 x 0.05 = 0.0019. Prescribing treatment B does not increase patient mortality rate as much as prescribing standard care and the expected loss reflects that. Treatments C and D will be prescribed after fewer than 0.01% trials and the expected loss from these errors is negligible. Overall expected loss in this scenario is 0.0275, 0.0257 resulting from prescribing standard care and 0.0019 from prescribing treatment B. Although standard care is only the third-best option, it is prescribed much more frequently than the second-best option (B) due to the status-quo deference inherent in hypothesis testing.

Panel B shows what would happen if the empirical success rule were used to prescribe treatments based on trial results. Treatment A would be prescribed after 93% of trials. The second-best treatment B would be prescribed after 7% of trials, resulting in expected loss of 0.07 x 0.05 = 0.0035. Standard care would be prescribed only after 0.02% of trials, and treatments C and D after fewer than 0.01% of trials. The overall expected loss when using the empirical success rule in this scenario is 0.0035.

Near-optimality is measured by considering all possible scenarios for the average outcomes of treatments in the trial. The Technical Appendix describes the algorithm used to compute near-optimality. In Table 4 we compare near-optimality of prescribing treatments using standard multiple hypothesis testing approach and of prescribing them using the empirical success rule in five-arm trials with different sample sizes. We report results both for trials with a 2:1:1:1:1 treatment-assignment ratio (as in the Recovery Trial) and for trials with the same total sample size, but balanced assignment of patients to treatments. In each case considered, the empirical success rule is more than 3 times nearer to optimality than the test-based decision criterion.

5. Near-Optimality of the Empirical Success Rule with Patient-Specific Treatment and Multiple Outcomes

The numerical calculations of near-optimality presented in Tables 1 through 4 concern relatively simple settings where patients are observationally identical and trial outcomes are binary, such as mortality. In clinical practice, trial outcomes may take multiple values. For example, trials of experimental drugs for COVID-19 may report both mortality outcomes and time to recovery for patients who survive. Patients who vary in age, gender, and comorbidities may vary in their response to treatment.

It has been common in analysis of trial data to designate primary and secondary outcomes. The latter are often called side effects. Research articles focus attention on the primary outcome. This is reasonable when the primary outcome is the dominant determinant of patient welfare or, put another way, when there is little variation in secondary outcomes across treatments. It is not reasonable otherwise. When the secondary effects of treatments are serious, it is more reasonable to consider how the primary and secondary outcomes jointly determine patient welfare. This is easy to accomplish with the empirical success rule. Box 2 uses the protocol for the Recovery Trial[5] to illustrate.

Methodological research has shown how to compute or bound the near-optimality of the empirical success rule when applied in a broad range of settings. We summarize the findings below.

---

[5] https://www.recoverytrial.net/files/recovery-protocol-v5-0-2020-04-24.pdf , accessed May 21, 2020.

---

Box 2: Possible Application of the Empirical Success Rule to Data from the Recovery Trial

The protocol for the Recovery Trial states "For each pairwise comparison with the 'no additional treatment' arm, the **primary objective** is to provide reliable estimates of the effect of study treatments on all-cause mortality at 28 days after first randomisation . . . . The **secondary objectives** are to assess the effects of study treatments on duration of hospital stay; the need for . . . . ventilation; and the need for renal replacement therapy. Data from routine healthcare records . . . . and from relevant research studies . . . . will allow subsidiary analyses of the effect of the study treatments on particular non-fatal events . . . . the influence of pre-existing major co-morbidity . . . . and longer-term outcomes . . . . as well as in particular sub-categories of patient . . . . ."

Thus, the researchers intend to study multiple outcomes and to examine heterogeneity of treatment response across patients who vary in co-morbidities and other covariates. The analysis plan calls for pairwise comparison of each experimental treatment with 'no additional treatment;' that is, with standard care. The plan refers to use of hypothesis tests to make these comparisons, particularly the Dunnett test for multiple comparisons.

We suggest application of the empirical success rule. A simple approach would be to model overall patient welfare as a weighted sum of the multiple outcomes described in the protocol, each outcome weighted appropriately to reflect its contribution to welfare. A clinician might, for example, find it desirable to give large positive weight to survival and some negative weight to duration of hospital stay, the need for ventilation, and the need for renal replacement therapy. Considering patients who share similar observed comorbidities and other covariates, each treatment would be evaluated by its empirical success in achieving overall patient welfare in this covariate group.

---

Near-Optimality with Binary Primary and Secondary Outcomes

Manski and Tetenov (2019) study the near-optimality of the empirical success rule when there are two feasible treatments and patient welfare is a weighted sum of binary primary and secondary outcomes. The primary outcome is patient survival for a specified time period. The secondary one denotes whether the patient suffers a specified side effect of treatment.

When a patient does not suffer the side effect, we let welfare equal 1 if a patient survives and equal 0 if he does not survive. When a patient experiences the side effect, welfare is lowered by a specified fraction $h$, whose value expresses the harm associated with the side effect. Thus, a patient who experiences the side effect has welfare $1 - h$ if he survives and $-h$ if he does not survive.

In this setting, we develop an algorithm to compute the near-optimality of the empirical success rule, which evaluates trial data by the frequencies of survival and the side effect observed with each treatment. We present numerical findings for alternative values of sample size and the value h expressing the harm of the side effect.

Near-Optimality with Bounded Outcomes

Exact computation of near-optimality is feasible when trial outcomes are binary or take only a few values, but it becomes more onerous when outcomes can take many values or are continuous. When outcomes are bounded, large-deviations inequalities of probability theory yield upper bounds on the near-optimality of the empirical success rule. These upper bounds provide conservative measures of near-optimality. Their value is that they are simple to compute and are sufficiently informative to provide useful guidance to clinicians.

Research of this type was initiated by Manski (2004), who used a large-deviations inequality for sample averages of bounded outcomes to derive an upper bound on the near-optimality of the empirical success rule when used to choose between two treatments. Manski and Tetenov (2016) extend the analysis to multi-arm trials. Their Proposition 1 extends the early finding of Manski (2004) from two to multiple treatments. Proposition 2 derives a new large-deviations bound for multiple treatments.

Let $L$ be the number of treatment arms and let $V$ be the range of the bounded outcome. When the trial has a balanced design, with $n$ subjects per treatment arm, the upper bounds on near-optimality proved in Propositions 1 and 2 have particularly simple forms, being $(2e)^{-\frac{1}{2}}V(L-1)n^{-\frac{1}{2}}$ and $V(\ln L)^{\frac{1}{2}}n^{-\frac{1}{2}}$. The former result provides a tighter bound than the latter for two or three treatments, while the latter result gives a tighter bound for four or more treatments. In both cases, the upper bound decreases toward zero at rate √n as the number n of subjects per arm increases.

Near-Optimality with Heterogeneous Patients

Patient response to treatments for COVID-19 may be heterogenous, varying with observable covariates including age, gender, and comorbidities. Hence, a clinician may want to assess the near-optimality of a decision criterion when applied to patients who share similar observed covariates.

In principle, this is easy to do. The clinician may view each group of patients who share similar covariates as a separate patient population. Accordingly, the clinician may apply the empirical success rule separately to each group, choosing a treatment that yields the highest average outcome among the trial participants who have the group covariates. In this manner, patient care may recognize heterogeneity of treatment response.

In practice, the ability of clinicians to differentially treat patients with different covariates is sometimes limited by the failure of medical researchers to report how trial findings vary with patient covariates. A common rationale is concern with statistical significance. Stratifying trial participants into covariate groups usually reduces the statistical precision of estimates of treatment effects. Research articles often report only findings that are statistically significant by conventional criteria.

Information is lost when reporting research findings is tied to statistical significance. It is important to study and report observable heterogeneity in treatment response to the extent feasible. The analysis of this paper makes clear that estimates of treatment effects need not be statistically significant to be clinically useful.

6. Discussion

A central objective of empirical research on treatment response is to inform treatment choice. Yet researchers analyzing trial data have used concepts of statistical inference whose foundations are distant from treatment choice. It has been common to use hypothesis tests to choose treatments. We evaluate decision criteria by near-optimality and suggest this way to analyze findings of trials comparing COVID-19 treatments. From this perspective, the empirical success rule performs much better than hypothesis testing.

In contrast to hypothesis tests, the empirical success rule views treatments symmetrically. Of course, use of the rule does not guarantee that the optimal treatment is always chosen. No decision criterion can achieve this ideal with finite trial data. Evaluation of criteria by near-optimality appropriately recognizes how the probability and magnitude of errors in decision making combine to affect patient welfare.

For simplicity, we have considered randomized trials having full internal and external validity. Internal validity may be compromised by non-compliance and loss to follow up. External validity may be compromised by measurement of surrogate outcomes and by administration of trials to types of patients who differ from those that clinicians treat in practice. The concept of near-optimality is applicable when analyzing data from trials with limited validity, but the specific numerical calculations made in this paper would require modification.

A limitation of this paper is that it only considers treatment choice using data from one trial. In practice, a clinician may learn the findings of multiple trials and may also be informed by observational data. The concept of near-optimality is well-defined in these more complex settings, but methods for practical application are yet to be developed.

A further issue beyond the scope of this paper concerns the dynamics of treatment choice when new trials and observational evidence may emerge in the future. The concept of near-optimality should be extendable to such settings. However, methodology for application is yet to be developed.

Dynamic analysis of treatment choice made with hypothesis tests may be especially difficult to perform, because testing views standard care and new treatments asymmetrically. As new evidence accumulates over time, the consensus designation of standard care may change, leading to a change in the null hypothesis when new trials are evaluated. The implications for patient welfare are unclear.

Technical Appendix

General setup

We use concepts and notation like those in Manski (2004) and Manski and Tetenov (2016, 2019). The clinician must assign one of $L$ treatments studied in the clinical trial to each member of a treatment population, denoted $J$. Denote the set of treatments by $T = \{1, 2, \ldots, L\}$, treatment 1 being standard care. Each individual $j \in J$ has a response function $y_j(\cdot): T \to Y$ mapping treatments $t \in T$ into individual patient-relevant outcomes $y_j(t) \in Y$. In general, outcomes could be multi-valued and multi-dimensional. For example, the relevant outcomes for COVID-19 treatment may be survival, taking the value 0 or 1, and time to recovery for those who survive, measured in number of days.

The probability distribution $P[y(\cdot)]$ of the random function $y(\cdot): T \to Y$ describes treatment response across the population. The distribution $P$ is unknown. The set of all feasible distributions $P$ is $\{P_s, s \in S\}$, where $S$ indexes all feasible *states of nature*. When computing near-optimality in Tables 2 and 4, we include in $S$ all logically possible outcome distributions.

We assume that *patient welfare* is a known function $u$: $Y \rightarrow \mathbf{R}$ of individual outcomes. For binary outcomes $Y = \{0, 1\}$, with 1 denoting success. In this special case, it is without loss of generality to set $u(y) = y$. For two-dimensional patient outcomes $y = (y_p, y_{se})$, where $y_p$ denotes the primary outcome and $y_{se}$ the side effect, Manski and Tetenov (2019) considered patient welfare that is a weighted sum of the two outcomes: $u(y) = y_p - hy_{se}$, where $h$ expresses the harm of the side effect relative to the primary outcome.

Now consider data generation. Let $\Psi$ denote the sample space; that is, $\Psi$ is the set of data samples that could be generated by the trial. Let $Q_s$ denote the sampling distribution on $\Psi$ in state of nature $s$. That is, $Q_s$ is the probability distribution of different trial outcomes.

We consider trials that randomize a predetermined number of subjects $n_t$ to each treatment $t$. The set $n_T \equiv [n_t, t \in T]$ of stratum sample sizes defines the design. The total number of subjects in the trial is then $N \equiv \sum_{t \in T} n_t$. The data $\psi$ are the $N$ pairs of individual treatment assignments $t_i$ and outcomes $y_i$: $\psi = [(t_i, y_i), i = 1, 2, \ldots, N]$.

The sampling distribution $Q_s$ is determined by the probability distribution of treatment response $P_s$ and the trial design, with $Q_s(y_i|t_i) = P_s(y(t_i))$. We assume that treatment response is individualistic; that is, patient outcomes are statistically independent of the outcomes of other patients in the trial.

A statistical treatment rule maps sample data into a treatment allocation. A feasible treatment rule is a function that randomly allocates persons across the different treatments. Let $\Delta$ now denote the space of functions that map $T$ into the unit interval and that satisfy the adding-up condition: $\delta \in \Delta \implies \sum_{t \in T} \delta(t, \psi) = 1$, $\forall \psi \in \Psi$. Then each function $\delta \in \Delta$ defines a statistical treatment rule.

The mean welfare outcome of treatment $t$ in state of nature $s$ is denoted by $\mu_{st} \equiv \mathrm{E}_s[u(y(t))]$. The maximum average patient welfare achievable in state $s$ is $\max_{t \in T} \mu_{st}$. After trial data $\psi$ are observed, the fraction $\delta(t, \psi)$ of patients will be treated with treatment $t$, resulting in mean patient welfare $\sum_{t \in T} (\mu_{st} \delta(t, \psi))$. The mean welfare of patients treated according to statistical treatment rule $\delta$ over repeated realizations of the trial is then $\int_{\Psi} \sum_{t \in T} (\mu_{st} \delta(t, \psi)) \, dQ_s(\psi) = \sum_{t \in T} \mu_{st} E_s[\delta(t, \psi)]$, where $E_s[\delta(t, \psi)] = \int_{\Psi} \delta(t, \psi) dQ_s(\psi)$ is the expected (across potential samples) fraction of persons who will be assigned to treatment $t$.

Application of statistical treatment rule $\delta$ in state of nature $s$ leads to an expected loss (regret) equal to

(A1)  $\max_{t \in T} \mu_{st} - \sum_{t \in T} \mu_{st} E_s[\delta(t, \psi)]$.

The near-optimality (maximum regret) of statistical treatment rule $\delta$ is the maximum value of (A1) over all feasible states of nature:

(A2)  $\max_{s \in S} \left( \max_{t \in T} \mu_{st} - \sum_{t \in T} \mu_{st} E_s[\delta(t, \psi)] \right)$.

Hypothesis Testing Rules

First, we consider statistical treatment rules based on hypothesis tests for univariate outcomes $y$. Denote the sample mean of $y$ observed in arm $t$ of the trial by $\bar{y}_t = \frac{1}{n_t} \sum_{i: t_i = t} y_i$. To test the null hypothesis that all treatments have the same outcome distribution, we use $\hat{\sigma}^2 = \frac{1}{N-L} \sum_{t \in T} \sum_{i: t_i = t} (y_i - \bar{y}_t)^2$ as the estimator of common variance. Then the t-statistic for comparing the mean outcome of treatment $t = 2, \ldots, L$ with that of standard care (treatment 1) equals $\tau_t = \frac{\bar{y}_t - \bar{y}_1}{\hat{\sigma} \sqrt{1/n_t + 1/n_1}}$. Let $c$ be the critical value adjusted for multiplicity. Specifically, we use the Student's t-distribution for two-arm trials and the Dunnett's test critical value for multiple comparisons for multi-arm trials.

The hypothesis test rule prescribes treatment 1 (standard care) to everyone if all t-statistics are below the critical value.:

$$\delta_H(1, \psi) \equiv 1\left\{\max_{t \in \{2,\dots,L\}} \tau_t \leq c\right\}.$$

If some t-statistics comparing treatments 2,…,$L$ to standard care exceed the critical value, these treatments are considered statistically significantly better than standard care. We assume that among these treatments the one with the largest mean outcome in the trial will be prescribed (with equal probability if there is a tie).

$$\delta_H(t, \psi) \equiv \frac{1\left\{\tau_t > c, \ \bar{y}_t = \max_{t' \in \{2,\dots,L\}} \bar{y}_{t'}\right\}}{\sum_{t' \in \{2,\dots,L\}} 1\left\{\tau_t > c, \ \bar{y}_t = \max_{t' \in \{2,\dots,L\}} \bar{y}_{t'}\right\}}.$$

When treatment arms 2,…,$L$ have equal sample sizes, as in our Table 4, the t-statistics $\tau_t$ have the same ranking as the sample means $\bar{y}_t$. Hence, prescribing the treatment with the largest mean outcome in the trial is equivalent in this case to prescribing the treatment with the largest t-statistic.

The Empirical Success Rule

Let $\bar{u}_t = \frac{1}{n_t} \sum_{i:t_i=t} u(y_i)$ denote the mean patient welfare observed in treatment arm $t = 1, 2, \dots, L$. The empirical success rule considers all treatments in the trial symmetrically and prescribes the treatment with the largest observed mean patient welfare. If there is a tie, all treatments with the largest observed mean patient welfare are prescribed with equal probability.

$$\delta_{ES}(t, \psi) \equiv \frac{1\left\{\bar{u}_t = \max_{t' \in \{1,\dots,L\}} \bar{u}_{t'}\right\}}{\sum_{t' \in \{1,\dots,L\}} 1\left\{\bar{u}_t = \max_{t' \in \{1,\dots,L\}} \bar{u}_{t'}\right\}}.$$

For binary outcomes, we take $u(y) = y$.

Computation of near-optimality for two-arm trials with binary outcomes

When computing the near-optimality results reported in Table 2, we consider the set of all possible distributions of binary outcomes with means $p_1 \equiv \mathrm{E}[y(1)]$, $p_2 \equiv \mathrm{E}[y(2)]$, $(p_1, p_2) \in [0, 1]^2$.

Let $m_1$ and $m_2$ denote the number of positive outcomes in each arm of the trial. For binary outcomes, $\psi = (m_1, m_2)$ is a sufficient statistic for the sample. Hence, it is sufficient to consider the sample space $\Psi = \{0, 1, \ldots, n_1\} \times \{0, 1, \ldots, n_2\}$. The probability density function of $\psi$ is a product of two binomial density functions. This sample space is sufficiently small, so we compute (A1) exactly.

The function (A1) is continuous in $(p_1, p_2)$ but may have multiple global and local maxima. We approximate the maximum in (A2) by grid search using 1000 possible values for each parameter equally spaced on [0,1]: $\{0.0005, 0.0015, \ldots, 0.9995\}$.

Computation of near-optimality for multi-arm trials with binary outcomes

To compute the results reported in Table 4, we consider the set of all possible distributions of binary outcomes with means $p_t \equiv \mathrm{E}[y(t)]$, $t = 1, \ldots, L$, $(p_1, \ldots, p_L) \in [0, 1]^L$. Let $m_t$ denote the number of positive outcomes in arm $t$ of the trial. For binary outcomes, $\psi = (m_1, \ldots, m_L)$ is a sufficient statistic for the sample. Hence, we consider the sample space $\Psi = \{0, 1, \ldots, n_1\} \times \ldots \times \{0, 1, \ldots, n_L\}$. The large size of the sample space makes it impractical to evaluate (A1) exactly. Instead, given each value of $(p_1, \ldots, p_L)$ we simulate a large number of trial outcomes to approximate the sampling distribution $Q_s$. Our computations of the maximum of (A2) proceed in three steps.

First, we conduct a grid search using 51 possible values for each parameter $p_t \in [0, 0.02, \ldots, 1]$. For each combination of parameters, we approximate the sampling distribution $Q_s$ by simulating 100,000 trial outcomes. The results of this grid search suggest that the largest expected loss for the empirical success rule occurs when the parameters have the form $p_1 = a$, $p_2 = p_3 = p_4 = p_5 = b$, $a > b$. The largest expected loss for the Dunnett's test rule occurs when $p_1 = a$, $p_2 = b$, $p_3 = p_4 = p_5 = c$, $b > a$, $b > c$.

In the second step, we conduct a grid search over these two lower-dimensional parameter spaces using 101 possible parameter values from [0, 0.01, …, 1] for $a$, $b$, and $c$. In this step we approximate $Q_s$ by simulating 1,000,000 trial outcomes.

In the last step, we take 10 parameter combinations yielding the largest estimated expected loss for each decision rule in step 2 and re-compute expected loss by simulating 100,000,000 trial outcomes. We do this to verify that our results are not affected by bias resulting from approximating $Q_s$ by simulation.

The MATLAB code used to perform the computations is available from Aleksey Tetenov.

| Mortality rates: | | | | | | | |
|---|---|---|---|---|---|---|---|
| standard care alone | 0.25 | 0.25 | 0.25 | 0.25 | 0.25 | 0.25 | 0.25 |
| with new treatment | 0.4 | 0.35 | 0.3 | 0.25 | 0.2 | 0.15 | 0.1 |
| **Panel A: What happens if treatment decisions are made using a two-sided 5% hypothesis test** | | | | | | | |
| % of trials after which standard care will be prescribed | 100.00% | 99.98% | 99.70% | 97.50% | 86.76% | 57.36% | 18.92% |
| Loss from prescribing standard care | 0 | 0 | 0 | 0 | 0.05 | 0.1 | 0.15 |
| % of trials after which new treatment will be prescribed | 0.00% | 0.02% | 0.30% | 2.50% | 13.24% | 42.64% | 81.08% |
| Loss from prescribing new treatment | 0.15 | 0.1 | 0.05 | 0 | 0 | 0 | 0 |
| Expected loss: | 0.0000 | 0.0000 | 0.0002 | 0.0000 | 0.0434 | 0.0574 | 0.0284 |
| **Panel B: What happens if treatment decisions are made using the empirical success rule** | | | | | | | |
| % of trials after which standard care will be prescribed | 98.95% | 94.28% | 79.61% | 51.64% | 21.18% | 4.22% | 0.26% |
| Loss from prescribing standard care | 0 | 0 | 0 | 0 | 0.05 | 0.1 | 0.15 |
| % of trials after which new treatment will be prescribed | 1.05% | 5.72% | 20.39% | 48.36% | 78.82% | 95.78% | 99.74% |
| Loss from prescribing new treatment | 0.15 | 0.1 | 0.05 | 0 | 0 | 0 | 0 |
| Expected loss: | 0.0016 | 0.0057 | 0.0102 | 0.0000 | 0.0106 | 0.0042 | 0.0004 |

Table 1: Illustrative scenarios for a trial assigning 100 patients to standard care and 99 to a new treatment, as in Cao *et al.* (2020).

| Sample size per arm: | Near-optimality if treatment decisions are made using a two-sided 5% hypothesis test | Near-optimality if treatment decisions are made using the empirical success rule |
|---|---|---|
| 20 | 0.1685 | 0.0269 |
| 30 | 0.1304 | 0.0220 |
| 50 | 0.0990 | 0.0170 |
| 100 | 0.0705 | 0.0120 |
| 200 | 0.0510 | 0.0085 |
| 500 | 0.0319 | 0.0054 |
| 1000 | 0.0228 | 0.0038 |
| 2000 | 0.0161 | 0.0027 |
| 4000 | 0.0115 | 0.0019 |
| 5000 | 0.0102 | 0.0017 |
| 10000 | 0.0073 | 0.0012 |
| 15000 | 0.0059 | 0.0010 |

Table 2: Near-optimality of hypothesis test and empirical success decision rules for two-arm trials with equal number of patients in each arm.

| | Standard care | A | B | C | D |
|---|---|---|---|---|---|
| Sample size in each arm | 500 | 250 | 250 | 250 | 250 |
| Mortality rate of each treatment | 0.25 | 0.15 | 0.20 | 0.30 | 0.35 |
| Panel A: What happens if treatment decisions are made using two-sided Dunnett's test at 5% significance | | | | | |
| % of trials after which new treatment will be prescribed | 25.65% | 70.60% | 3.75% | 0 | 0 |
| Loss from prescribing each treatment | 0.1 | 0 | 0.05 | 0.15 | 0.2 |
| Probability of error times the magnitude of loss | 0.0257 | 0 | 0.0019 | 0 | 0 |
| Expected loss given these mortality rates | | | | | **0.0275** |
| Panel B: What happens if treatment decisions are made using the empirical success rule | | | | | |
| % of trials after which new treatment will be prescribed | 0.02% | 92.95% | 7.03% | 0 | 0 |
| Loss from prescribing each treatment | 0.1 | 0 | 0.05 | 0.15 | 0.2 |
| Probability of error times the magnitude of loss | 0 | 0 | 0.0035 | 0 | 0 |
| Expected loss given these mortality rates | | | | | **0.0035** |

Table 3: Illustrative scenario for a multi-arm clinical trial assigning 500 patients to receive standard care and 250 patients each to four alternative treatments.

| Sample sizes for each arm: | Near-optimality if treatment decisions are made using a two-sided 5% Dunnett's test | Near-optimality if treatment decisions are made using the empirical success rule |
|---|---|---|
| 100:50:50:50:50 | 0.1224 | 0.0362 |
| 60:60:60:60:60 | 0.1251 | 0.0343 |
| 200:100:100:100:100 | 0.0855 | 0.0256 |
| 120:120:120:120:120 | 0.0859 | 0.0243 |
| 500:250:250:250:250 | 0.0532 | 0.0160 |
| 300:300:300:300:300 | 0.0563 | 0.0153 |
| 1000:500:500:500:500 | 0.0380 | 0.0112 |
| 600:600:600:600:600 | 0.0390 | 0.0107 |
| 2000:1000:1000:1000:1000 | 0.0274 | 0.0080 |
| 1200:1200:1200:1200:1200 | 0.0291 | 0.0076 |

Table 4: Near-optimality of multiple hypothesis testing and empirical success decision rules for five-arm trials with specified sample sizes.